\documentclass[%
reprint,
superscriptaddress,
 amsmath,amssymb,
 aps,
 pre,
]{revtex4-2}

\usepackage{CJK}
\usepackage{graphicx}%
\usepackage{dcolumn}%
\usepackage{bm}%
\usepackage[breaklinks=true,
            colorlinks=true,
            linkcolor=blue,
            citecolor=cyan]{hyperref} %

\usepackage{soul} %
\usepackage[dvipsnames]{xcolor}

\begin{document}

\begin{CJK*}{UTF8}{gbsn}

\title{Perfect adaptation in eukaryotic gradient sensing\\
using cooperative allosteric binding}%

\author{Vishnu Srinivasan}
\affiliation{%
Department of Physics and Astronomy, Johns Hopkins University, Baltimore, Maryland 21218, USA
}%

\author{Wei Wang (汪巍)}
\affiliation{%
Department of Physics and Astronomy, Johns Hopkins University, Baltimore, Maryland 21218, USA
}%

\author{Brian A. Camley}%
\affiliation{%
Department of Physics and Astronomy, Johns Hopkins University, Baltimore, Maryland 21218, USA
}%
\affiliation{%
Department of Biophysics, Johns Hopkins University, Baltimore, Maryland 21218, USA
}

\begin{abstract}
Eukaryotic cells generally sense chemical gradients using the binding of chemical ligands to membrane receptors. In order to perform chemotaxis effectively in different environments, cells need to adapt to different concentrations. We present a model of gradient sensing where the affinity of receptor-ligand binding is increased when a protein binds to the receptor's cytosolic side. This interior protein (allosteric factor) alters the sensitivity of the cell, allowing the cell to adapt to different ligand concentrations. We propose a reaction scheme where the cell alters the allosteric factor's availability to adapt the average fraction of bound receptors to $1/2$. We calculate bounds on the chemotactic accuracy of the cell, and find that the cell can reach near-optimal chemotaxis over a broad range of concentrations. We find that the accuracy of chemotaxis depends strongly on the diffusion of the allosteric compound relative to other reaction rates. From this, we also find a trade-off between adaptation time and gradient sensing accuracy.
\end{abstract}

\maketitle
\end{CJK*}

\section{Introduction}\label{sec:intro}
Eukaryotic cells can sense and follow chemical gradients (a process called ``chemotaxis''); this is essential to processes like wound healing, immune response, and cancer  metastasis \cite{jin2008chemotaxis, roussos2011cancer,sengupta2021principles}. Eukaryotic chemotaxis has been shown to have remarkable sensitivity to shallow chemical gradients, with some studies finding that cells can sense gradients on the order of 1--2\%~\cite{song2006dictyostelium, ueda2007stochastic} over a cell width of a few microns. In eukaryotes, chemotaxis occurs by cells recording concentration measurements at different points in space, using membrane receptors to which ligands in the surrounding fluid are bound, creating downstream signals of cell polarity and motility ~\cite{swaney2010eukaryotic}. Since the biochemistry of ligand-receptor binding is subject to noise via thermal processes, the resulting signal is stochastic in nature~\cite{bialek2012biophysics,ueda2007stochastic,levine2013physics}. %
In many circumstances, if the stochasticity arising from ligand-receptor binding limits the ability of a cell to sense a gradient, gradient sensing accuracy is optimal in a narrow concentration range near the ligand-receptor dissociation constant \cite{hu2010mle,ueda2007stochastic,hu2011geometry,segota2013high,lauffenburger1982influence}. This narrow range may be a problem for cells whose natural environments are challenged with a broad range of chemoattractant concentrations \cite{kashyap2024trade}. 
Is there a mechanism where cells can keep their sensing accuracy over wide concentration ranges? One possibility, motivated by the observation of multiple receptor types for a single chemoattractant \cite{dewit1992multiple, benshlomo2005}, is that cells may hedge their bets by expressing multiple receptor types with different dissociation constants \cite{hopkins2020chemotaxis} but in doing so sacrifice peak sensing accuracy. Another possibility, which we study here, is an adaptive modification to receptors to maintain chemotactic accuracy. Receptor-level adaptation is well-known in bacterial systems but not believed to be present in eukaryotes \cite{tu2018adaptation} -- eukaryotic chemotaxis has adaptation mechanisms downstream of receptor binding \cite{takeda2012incoherent}. 
In this paper, we propose a model that allows eukaryotic cells to effectively adapt ligand-receptor affinity through the regulation of a protein that allosterically changes the dissociation constant of the receptor. This could include a G protein, arrestin, or any allosteric modulator well described by a ternary complex model \cite{may2003allosteric,carvalho2021novel,de1980ternary,thal2018structural}. If the allosteric protein's availability is regulated by the local fraction of receptors bound, the cell can control dissociation constants to ensure the fraction bound is close to 1/2, where chemotaxis is most accurate. We find that near-perfect adaptation is possible under certain constraints on the parameters involved (Secs.~\ref{sec:model-ternary}--\ref{sec:adaptation-diffusion}). We claim that the diffusion coefficient of this allosteric protein relative to its activation rate, is what decides whether such adaptation is advantageous for the cell, which we evaluate by looking at the gain in accuracy over simple single-receptor-type chemotaxis or bet-hedging. However, if the activation rate is slow, allowing near-perfect adaptation, this comes with an unavoidable tradeoff of a long time required to adapt to new concentrations. This may explain why, even though eukaryotic cells have all the necessary requirements to perfectly adapt at the receptor level, no evidence of this has been seen. %

\section{Model and results}

We review how a cell's ability to sense a chemical gradient is related to the stochasticity of its receptor-ligand interactions in Sec.~\ref{sec:model-MLE}. We then use these results to understand how gradient sensing is modified depending on the concentration of an allosteric protein in a model where it can enhance ligand-receptor binding in Sec.~\ref{sec:model-ternary}. We then propose in Sec.~\ref{sec:model-MM} a scheme where cells can adapt their concentration of active allosteric proteins to maximize the cell's ability to sense a chemical gradient's orientation. In Sec.~\ref{sec:adaptation-diffusion} we show this scheme requires sufficiently fast diffusion of allosteric protein to lead to perfect adaptation, and show in Sec.~\ref{sec:tradeoffs} that this constraint implies a tradeoff between accuracy of sensing and time required to adapt to new concentration levels. 

\subsection{Gradient sensing accuracy depends on ligand binding probabilities}\label{sec:model-MLE}

We model cells in two dimensions as circular with receptors on the boundary. We assume cells are in an exponential concentration gradient,
\begin{equation}\label{eq:concentration-profile}
    C(\theta) = C_0\exp\left[\frac{g}{2} \cos(\theta-\phi)\right],
\end{equation}
where $C_0$ is the mean concentration and $g$ the percentage change across the cell, $R$ is the cell radius, %
$\theta$ is the angular coordinate of a receptor and $\phi$ is the gradient direction. {We will later assume shallow gradients $g \ll 1$ to make certain analytical calculations tractable.}  
The state of the $i$th receptor can be modeled as a Bernoulli random variable  $X_i$, where $X_i=1$ indicates the receptor is bound and $X_i=0$ indicates the receptor is unbound. 
 If there are $N$ receptors on the cell membrane, the associated likelihood function is:
\begin{equation}\label{eq:likelihood-gen}
    \mathcal{L}\left(g, \phi|\{X_i\}\right) = \prod\limits_{i=1}^N  p_{\textrm{bound},i}^{X_i}(1-p_{\textrm{bound},i})^{1-X_i}.
\end{equation}
We can then extract the Cram\'er-Rao bound for the variance of the unbiased estimators $\hat{\bm{\theta}} \equiv (\hat{g},\hat{\phi})$ of maximum likelihood estimation (MLE)~\cite{hu2011geometry, wang2024limits}:
\begin{equation}
    {\textrm{Var}(\hat{\theta}_{\mu})}\geqslant [I(\bm{\theta})^{-1}]_{\mu\mu},
\end{equation}
where $I(\bm{\theta})^{-1}$ denotes the inverse of the Fisher information matrix $I_{\mu\nu} = -\langle\partial^2\ln\mathcal{L}/\partial\theta_\mu\partial\theta_\nu\rangle$, {where the indices $\mu,\nu$ run over the gradient magnitude $g$ and orientation $\phi$.}
This gives a lower bound for the variance of {\it any} unbiased
estimator -- this reflects the best estimate that a cell can make given the information it has. 
Hu \textit{et al.}~\cite{hu2011geometry} calculated these bounds for a 2D circular cell with the concentration profile of Eq.~\eqref{eq:concentration-profile}, and assuming that there are simple ligand-receptor kinetics. In this case, the probability of being bound is $p_{\textrm{bound},i} = {C(\theta_i)}/{\left[C(\theta_i)+K_D\right]}$, where $K_D$ is the ligand-receptor dissociation constant. %
The $(\phi, \phi)$ term in the Fisher information matrix is then \cite{hu2011geometry}
\begin{equation}\label{eq:FI-single-receptor}
    I_{\phi\phi} = \frac{Ng^2C_0K_D}{8(C_0+K_D)^2}.
\end{equation}
We think of $I_{\phi\phi}$ as characterizing the minimal error possible in a cell's estimation of the direction of the gradient -- the Cram\'er-Rao bound for the gradient direction is $\textrm{Var}(\hat{\phi})\geqslant I^{-1}_{\phi\phi}$. Sweeping over the mean concentration $C_0$, the Fisher Information peaks at $C_0=K_D$, and falls off monotonically on either side of the peak. 

We can generalize Eq.~\eqref{eq:FI-single-receptor} to an arbitrary shallow-gradient concentration profile by expanding $p_\textrm{bound}$ in terms of $g$. Given a bound probability $p_{\textrm{bound}}(\theta) = \xi_0 + \xi_1g\cos(\theta-\phi) + \mathcal{O}(g^2)$, we find $I_{\phi\phi}$ as
\begin{equation}\label{eq:FI-pbound}
    I_{\phi\phi}  = \frac{N\xi_1^2g^2}{2\xi_0(1-\xi_0)} + \mathcal{O}(g^3).
\end{equation}
The details of this calculation can be found in Appendix~\ref{app:fisher-information}. This proves to be a useful result as it allows for a quick evaluation of gradient sensing accuracy given the spatial profile of receptor binding.

\begin{figure}
    \includegraphics[width=0.7\linewidth]{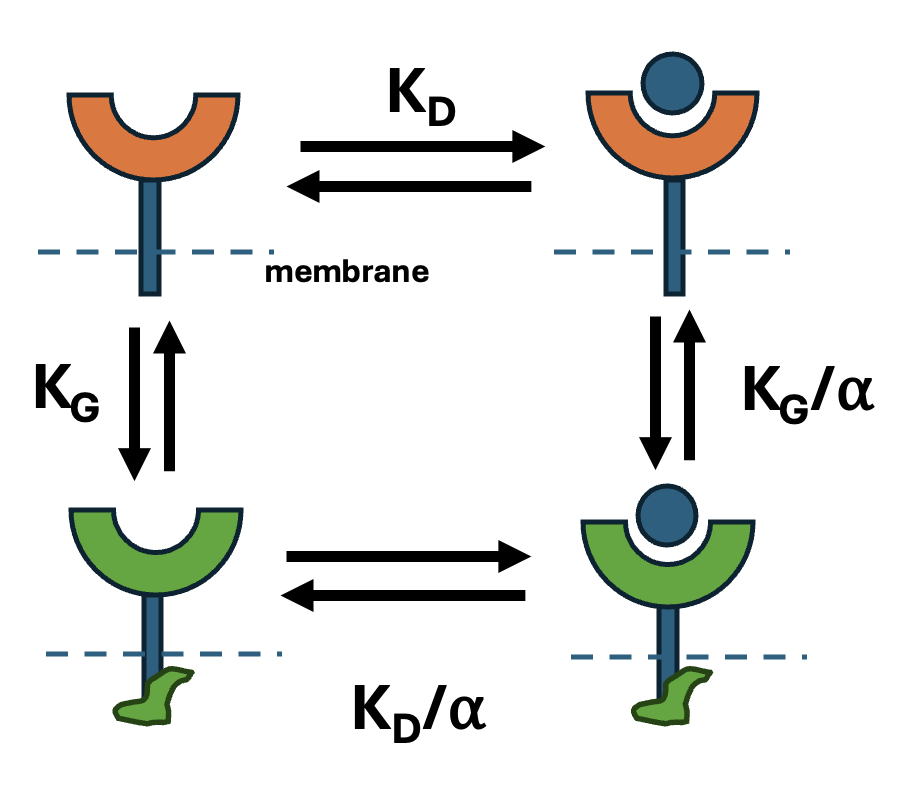}
    \caption{Reaction diagram showing all four states. Blue dots are ligand molecules, and green shape is the allosteric protein molecule bound to receptor. Labels on double-ended arrows are dissociation constants associated with each pair of states.} %
    \label{fig:ternary-complex}
\end{figure}

\subsection{Cooperative ternary interactions allow the cell to tune its region of maximal accuracy}\label{sec:model-ternary}
We model the allosteric protein-receptor interaction using a ternary complex model~\cite{de1980ternary} involving receptor, ligand and allosteric protein. Each receptor can bind one ligand molecule on the extracellular side, and one protein molecule on the cytosolic side (Fig.~\ref{fig:ternary-complex}). The dissociation constants for these processes  are $K_D$ for ligand binding to the bare receptor and $K_G$ for the allosteric protein binding to the bare receptor. When this protein binds to the receptor, this decreases the ligand-receptor dissociation constant from $K_D $ to $K_D/\alpha$. %
Unless otherwise noted, we assume $\alpha>1$, i.e., cooperative binding -- the ligand-receptor interaction is stronger if the G protein is bound -- and take $\alpha$ to be of order 10~\cite{de1980ternary}. 
 From detailed balance, the dissociation constant associated with the allosteric protein binding must be $K_G/\alpha$ when a ligand is bound~\cite{pitaevskii2012physical}.
In this model, the total probability that a receptor is bound to a ligand molecule can be written as 
\begin{equation}\label{eq:p_bound}
    p_{\text{bound}} = \frac{C}{C+K_{\text{eff}}},
\end{equation}
with an effective dissociation $K_{\text{eff}}$ given by
\begin{equation}\label{eq:K_eff}
    K_{\text{eff}} = K_D \frac{G + K_G}{\alpha G + K_G}, 
\end{equation}
where $G$ is the allosteric protein concentration (Appendix \ref{app:ternary-complex-sol}).

Eq.~\eqref{eq:K_eff} shows that the effective ligand-receptor interaction can have a dissociation ranging from $K_D/\alpha$ (in the limit $G \gg K_G$) to $K_D$ (in the limit $G \ll K_G$). If all receptors on the membrane have bound probability of Eq.~\eqref{eq:p_bound}, we can compute the Fisher information from Eq.~\eqref{eq:FI-pbound}, finding: 
\begin{equation}\label{eq:FI-Keff}
    I_{\phi\phi} = \frac{Ng^2C_0K_\textrm{eff}}{8(C_0+K_\textrm{eff})^2}.
\end{equation}
{We find that using the ternary complex model, we get a Fisher information that is exactly that of Eq.~\eqref{eq:FI-single-receptor} but with the effective dissociation constant $K_\textrm{eff}$}. The Fisher information $I_{\phi\phi}$ is thus maximized when $C_0 = K_\text{eff}$. This point of maximal accuracy is when the probability of binding is 1/2, by Eq.~\eqref{eq:p_bound}. The presence of the allosteric protein allows the cell to \emph{tune} its region of peak efficiency to any value between $K_D/\alpha$ and $K_D$ by changing $G$. We show that this can lead to perfect adaptation of sensing accuracy below in Sec.~\ref{sec:model-MM}.

\begin{figure}
    \includegraphics[width=0.43\textwidth]{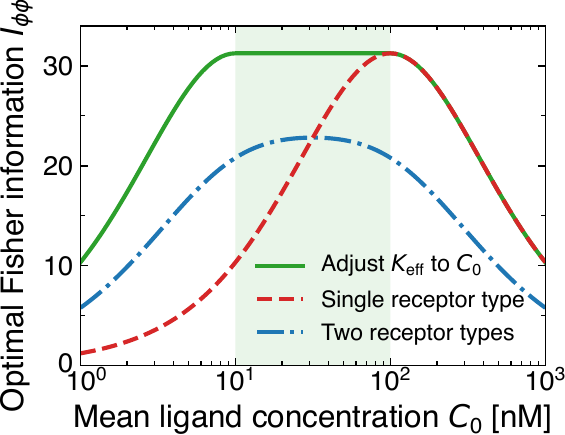}
    \caption{  
    How the Fisher information $I_{\phi\phi}$ varies with the mean ligand concentration $C_0$.
    The green solid line represents the case in which all receptors adapt their effective dissociation constant $K_\textrm{eff}$ to match $C_0$; the light green shaded region indicates the range over which perfect adaptation, $K_\textrm{eff} = C_0$, is achievable, i.e., $C_0\in[K_D/\alpha, K_D]$. For comparison, the results for the single-receptor-type model (red dashed line) and the two-receptor-type model (blue dash-dotted line) are also shown, indicating that the adaptation model is the optimal model among the three. Dissociation constant $K_D=100\,$nM and $\alpha=10$.}
    \label{fig:F_comparing}
\end{figure}

If $K_\textrm{eff}$ takes on its optimal possible value, i.e. $K_\textrm{eff} = C_0$ when $K_D/\alpha \leq C_0 \leq K_D$, we find 
\begin{equation}\label{eq:FI-optimal}
    I^\textrm{opt}_{\phi\phi} = \frac{N g^2}{8} \begin{cases} 
        \dfrac{C_0K_D / \alpha}{(C_0+K_D/\alpha)^2}, & C_0\leq K_D/\alpha,\\
      1/4, & K_D/\alpha \leq C_0 \leq K_D,\\
      \dfrac{C_0K_D }{(C_0+K_D)^2}, & C_0 \geq K_D.
   \end{cases}
\end{equation}
We plot this optimal Fisher information in Fig.~\ref{fig:F_comparing}. We see that the Fisher information is near its maximum value for a broad range of concentrations. 

We {compare the optimal adaptation result of Eq. \eqref{eq:FI-optimal}} to two possible alternatives. First, a cell with a simple single receptor with dissociation constant $K_D$, leads to a clear peak in information near $K_D$. Secondly, if instead the cell has $N/2$ receptors with dissociation constant $K_D$ and $N/2$ with $K_D/\alpha$, this hedges the cell's bets~\cite{hopkins2020chemotaxis}---this allows cells to chemotax accurately at a broader range of concentrations but at a cost of lowering the maximum Fisher information. For example, a cell with two receptor types with dissociation constants $K_D$ and $K_D/\alpha$ has Fisher jnformation 
\begin{equation}\label{eq:FI-twotype}
    I_{\phi\phi}^\text{2-type} = \frac{N g^2}{16}\left(\frac{C_0K_D/\alpha}{(C_0+K_D/\alpha)^2} + \frac{C_0K_D}{(C_0+K_D)^2}\right).
\end{equation}
If a cell can actually reach the optimal adaptation of Eq.~\eqref{eq:FI-optimal}, it will have more information than either of these two alternatives. %

There is an implicit assumption behind Eq.~\eqref{eq:FI-Keff}. In calculating this, we assumed that the cell only ``knows'' whether or not a receptor is bound, i.e. that from the perspective of the cell, it cannot distinguish between the right two states in Fig.~\ref{fig:ternary-complex}). This is naturally an oversimplification of GPCR signaling. {For instance, if our allosteric protein were a G protein, in the canonical model of G protein-based signaling, activated G proteins must unbind from the receptor to lead to downstream signaling, something we do not take into account here. Our assumption here is made in large part so that our model will limit back to \cite{hu2010mle,hu2011geometry}. Different choices for which states are active/signaling (i.e. which states the cell ``knows'' about) and which states the cell can distinguish (i.e., whether the likelihood includes a binary $X_i$ or a ternary one) can be made and will lead to different results.}

\begin{figure}
    \centering
    \includegraphics[width=0.75\linewidth]{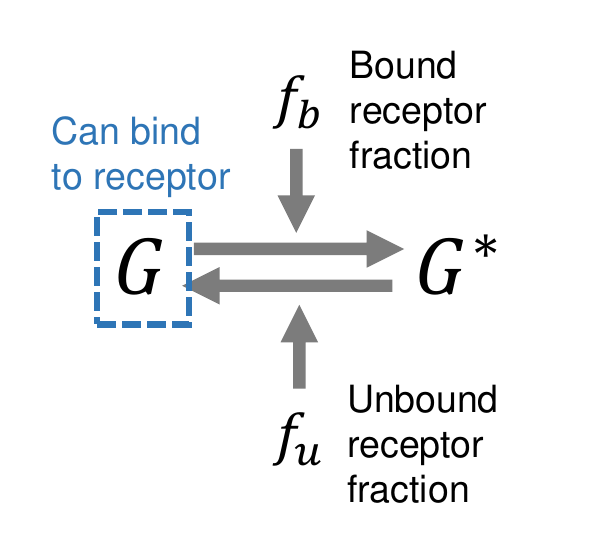}
    \caption{Reaction schematic for activation and inactivation of the allosteric protein in the form of \eqref{eq:mm-kinetics-nodiff}. A larger $f_u$ drives the activation of the allosteric protein, and a larger bound fraction $f_b$ drives its inactivation. Only the active form (labeled $G$) is able to bind to the receptor.}
    \label{fig:feedback_sketch}
\end{figure}

\begin{figure*}
    \includegraphics[width=0.95\textwidth]{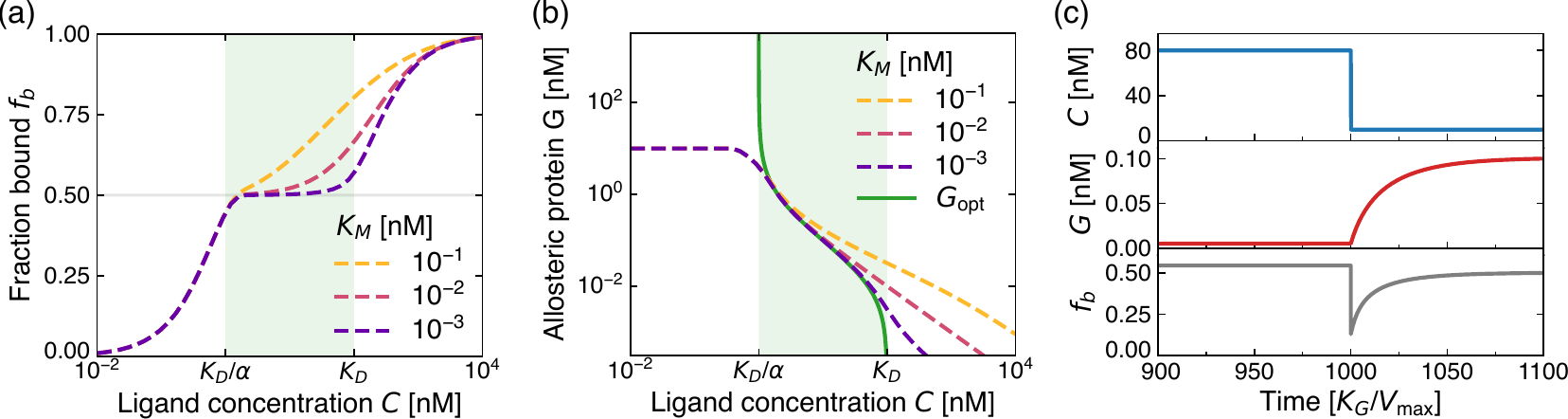}
    \caption{(a) Response curve illustrating the steady state receptor binding probability to a half-occupied state (gray line). Lower values of the Michaelis constant $K_M$ give better adaptation at a wider range of concentrations.
    Light green region indicates the range $C \in[K_D/\alpha,K_D]$, {where $\alpha=100$ is chosen to more clearly see the effect of $K_M$.}
    {(b) Corresponding active allosteric protein concentration $G$ in simulations for panel (a); green line indicates the optimal $G$ values from Eq.~\eqref{eq:K_eff} by setting $K_\mathrm{eff}=C$.}
    (c) Time-series response of the allosteric protein concentration $G$ (middle panel) and bound probability $p_\mathrm{bound}$ (bottom) in response to a sudden change in concentration from $80\,$nM to $10\,$nM (top). The value of $p_\mathrm{bound}$ initially decreases before the receptor adapts to the new concentration. We assume that ligand binding reaches steady state on a timescale much faster than that of adaptation, allowing us to treat the binding process as quasi-steady-state. Dissociation constant $K_D=100\,$nM and $\alpha=100$; Michaelis constant $K_M = 10^{-3}$ nM.
    \label{fig:KM_C0}}
\end{figure*}

\subsection{Feedback based on bound ligand fraction leads to perfect adaptation}\label{sec:model-MM}

For the cell to maximize its accuracy, it must change $G$ to tune $K_\textrm{eff}$ to the typical concentration of ligand around it, $C_0$ -- or to regulate the probability of binding to be 1/2. 
We propose a model where the allosteric protein switches between one state that is capable of binding to the receptor ($G$) and one that cannot ($G^*$) (Fig. \ref{fig:feedback_sketch}). {We assume that the reaction $G\to G^*$ is catalyzed either by the bound receptor or by an enzyme downstream of the bound receptor, so the rate is proportional to the bound receptor fraction, and similarly the reverse process catalyzed by unbound receptor is proportional to the unbound receptor fraction}. If these rates are of the Michaelis-Menten form \cite{johnson2011original,zim2025constraints}, then we can write an equation for the concentration of $G$ at a particular angular location $\theta$ along the cell:
\begin{equation}\label{eq:mm-kinetics-nodiff}
    \frac{\partial G}{\partial t} = V_{\text{max}}\left[-f_b(G,\theta)\frac{G}{G+K_M}  + f_u(G,\theta)\frac{G^*}{G^*+K_M}\right],
\end{equation}
{where $K_M$ is a Michaelis constant}, $G$ and $G^*$ are the active and inactive allosteric protein concentrations, $f_b$ and $f_u = 1-f_b$ are the local fraction of bound and unbound receptors at $\theta$. {We note that, in principle, the fraction bound is a stochastic variable, reflecting ligand-receptor binding fluctuations. We use the $f_b$ language to emphasize this potential complication. However, we neglect this stochasticity, assuming the fraction bound is just $f_b(G,\theta) \approx \langle f_b(G,\theta)\rangle = p_\textrm{bound}(G,C(\theta))$. Essentially, when we write this equation, we are writing an equation for a region of angles near $\theta$, and assuming $G$ is well-mixed over this region. Using the deterministic model seems reasonable if the number of receptors in this segment of the cell membrane is large, or the dynamics of $G$ is slow with respect to the kinetics of binding and unbinding, leading to effective time-averaging. Our deterministic model might not be appropriate in the limit of small diffusion coefficients, when the angular range where $G$ can be treated as well-mixed will be small. However, since we find our adaptation mechanism fails in the limit of small $D$ even in the deterministic case, we do not address this further.} 
Because we assume only the active form of this protein can bind to receptors, $f_b$ depends on $G$ through Eq.~\eqref{eq:p_bound}. $V_\text{max}$ is the maximum rate at which the allosteric protein  can be activated/deactivated. The total concentration of these proteins $G_\textrm{tot}=G+G^*$ is fixed, and so we have a corresponding equation for $G^*$:
\begin{equation}
\nonumber    \frac{\partial G^*}{\partial t} = V_{\text{max}}\left[f_b(G,\theta)\frac{G}{G+K_M} - f_u(G,\theta)\frac{G^*}{G^*+K_M}\right],
\end{equation}

In the limit where $G, G^*\gg K_M$, Eq.~\eqref{eq:mm-kinetics-nodiff} reduces to $\frac{\partial G}{\partial t} = V_{\text{max}}\left[-f_b + f_u\right]$ -- so the system will reach a steady state where $f_u = f_b = 1/2$ -- leading to perfect adaptation \cite{ferrell2016perfect}. However, because $K_\textrm{eff}$ can only range from $K_D/\alpha$ to $K_D$, the adaptation will not be able to reach this steady state when the concentration $C$ is outside this range. We also see that perfect adaptation requires $K_M$ to be small. We show in Fig.~\ref{fig:KM_C0}(a) a plot of fraction bound [Eq.~\eqref{eq:p_bound}] as a function of ligand concentration, where we find $G$ {[Fig.~\ref{fig:KM_C0}(b)]} by simulating Eq.~\eqref{eq:mm-kinetics-nodiff} until reaching steady state at each concentration {(see Appendix \ref{app:simulated-mm})}. %
We see that in the range from $K_D/\alpha$ to $K_D$ that the fraction bound adapts to $\sim 1/2$, and this adaptation is more accurate for smaller Michaelis constant. Outside of the range $K_D/\alpha < C < K_D$, {the concentration of active protein $G$ cannot go above the total concentration $G_\textrm{tot}$ or below zero [Fig.~\ref{fig:KM_C0}(b)] and adaptation fails. In the large and small-concentration limits, the binding probability behaves simply as if the receptors had dissociation constant $K_D/\alpha$ for $C<K_D/\alpha$, or $K_D$, for $C>K_D$.}

We show how the allosteric protein concentration and binding probability adapt to a step change in concentration in Fig.~\ref{fig:KM_C0}(c). The bound probability takes a sudden dip as concentration decreases, but as the  concentration of the allosteric protein increases (therefore decreasing $K_\textrm{eff}$), the probability of a ligand being bound evolves to near 1/2 again. We note that adaptation of the fraction bound here does not quite reach 1/2, due to the finite value of the Michaelis constant $K_M$ (e.g. as seen in \cite{ferrell2016perfect} for generic negative feedback adaptation).

\subsection{Combining adaptation and protein diffusion}\label{sec:adaptation-diffusion}
In the previous section, we implicitly assumed that the allosteric protein $G$ was localized at a particular angular location $\theta$. %
Though this scheme allows for perfect adaptation of $p_\textrm{bound}$, it is not ideal for gradient sensing, because each receptor will then adapt to the local concentration, resulting in each receptor being bound with probability $\approx 1/2$ where $K_\text{eff}(\theta)\approx C(\theta)$. In that case, there is no gradient of bound receptors across the cell (though there will be a gradient of allosteric proteins). 
{To maximize the change in number of bound receptors across the cell,  $K_\text{eff}$ should be constant across the cell -- essentially  adapting to the concentration $C_0$ at the cell midpoint, instead of the local concentration.} This can happen if the allosteric protein can diffuse on the cell membrane.  We therefore incorporate a diffusion term into Eq.~\eqref{eq:mm-kinetics-nodiff}: %
\begin{equation}\label{eq:mm-kinetics-diff}
    \frac{\partial G}{\partial t} = D\nabla^2G + \mathcal{R}(G, G^\ast),
\end{equation}
where the reaction term $\mathcal{R}(G, G^\ast)$ is just the right-hand side of Eq.~\eqref{eq:mm-kinetics-nodiff}. $G^\ast$ then obeys the complementary equation $\partial G^\ast/\partial t = D\nabla^2G^\ast -\mathcal{R}(G, G^\ast)$, so the total allosteric protein (active+inactive) concentration ($\oint (G+G^\ast)\mathrm{d}\mathbf{x}$) is conserved.
Note that on the 1D surface of a 2D circular cell, $\nabla\equiv (1/R)\partial_\theta$. The adaptive process now has natural scales of distance ($R$) and time ($K_G/V_{\text{max}}$), so we rescale Eq.~\eqref{eq:mm-kinetics-diff} to make it dimensionless:
\begin{equation}\label{eq:mm-kinetics-dimless}
    \frac{\partial \tilde G}{\partial \tilde{t}} = \tilde{D}\frac{\partial^2\tilde{G}}{\partial \theta^2} -f_b(\tilde{G},\theta)\frac{\tilde{G}}{\tilde{G}+\tilde{K}_M}  + f_u(\tilde{G},\theta)\frac{\tilde{G^*}}{\tilde{G^*}+\tilde{K}_M}, 
\end{equation}
where $\tilde{G}=G/K_G$, $\tilde{K}_M=K_M/K_G$, $\tilde{t} = tV_\text{max}/K_G$ and $\tilde{D}=DK_G/V_{\text{max}}R^2$.
When diffusion is slow relative to the switching between states of the allosteric protein, $\tilde{D}\ll1$, we expect that the bound fraction will be approximately 1/2 everywhere --i.e. the cell has almost no information about the gradient direction. As we increase $\tilde{D}$, we expect the cell's gradient sensing to approach the optimal case described in Sec.~\ref{sec:model-ternary} -- in this limit, it is equivalent to all the receptors adapting $K_\text{eff}$ to be the mean concentration $C_0$. 

\begin{figure}
    \includegraphics[width=0.48\textwidth]{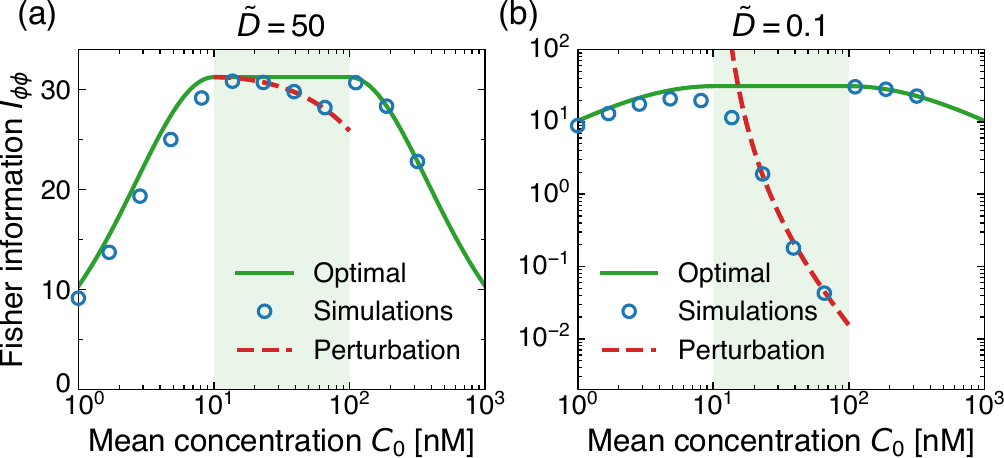}
    \caption{Comparing simulation results (blue circles) at high (a) and low (b) rescaled diffusivity $\tilde{D}$ with the analytical prediction (red dashed lines) from first-order perturbation theory [Eq.~\eqref{eq:perturbation-results}]. The green solid line from Fig.~\ref{fig:F_comparing}, corresponding to the limit $\tilde{D}\to\infty$, is included for reference. Dissociation constant $K_D=100\,$nM and $\alpha=10$.}
    \label{fig:checking perturbation}
\end{figure}

We solve Eq.~\eqref{eq:mm-kinetics-dimless} numerically, using forward time-centered space (FTCS) method with receptors on the surface of such a cell with radius $R$, using a time step $\Delta \tilde t=0.001$ and $\tilde K_M = 0.0001$ in the units described earlier. %
We simulate Eq.~\eqref{eq:mm-kinetics-dimless} until the equations reach their steady state value $G(\theta)$. Using these values of $G(\theta)$ then gives us a prediction for how the bound probability $p_\textrm{bound}$ depends on angle $\theta$. We fit this simulation result to the form $p_{\textrm{bound}}(\theta) = \xi_0 + \xi_1g\cos(\theta-\phi)$ using linear regression, which allows us to predict the Fisher information from Eq.~\eqref{eq:FI-pbound}. We can obtain steady-state Fisher information matrix elements for varying dimensionless diffusion coefficients $\tilde{D}$ (Fig.~\ref{fig:checking perturbation}). %
At large $\tilde{D}$, we see that cells with the adaptation process nearly reach the optimal possible Fisher information [Fig.~\ref{fig:checking perturbation}(a)], as we expect. However, we do note that even at $\tilde{D} = 50$ there is a slight decrease in $I_{\phi\phi}$ as %
 $C_0$ increases from $K_D/\alpha$ to $K_D$. By contrast, for low diffusion or fast reactions $\tilde{D} = 0.1$, the Fisher information can drop orders of magnitude below the optimal value, far worse than the alternatives shown in Fig. \ref{fig:F_comparing}. (We note that comparisons between the lowest values of the Fisher information should not be taken too seriously -- at very small Fisher information, the Cramer-Rao bound will be misleading because the variance of the angle $\phi$ is bounded because $\phi$ must be between $0$ and $2\pi$ \cite{nwogbaga2023physical,mardia2009directional}.) %

In addition to our numerical solutions to Eq.~\eqref{eq:mm-kinetics-dimless}, we use a perturbative approach to solve for the steady-state allosteric  protein concentration. We perturb about the $\tilde{D} = 0$ case to order $\tilde{D}$ and around $1/\tilde{D} = 0$ to order $1/\tilde{D}$. The full solution is somewhat lengthy and detailed in Appendix~\ref{app:perturbation-theory}. We obtain the following results for the perturbed Fisher information in the range $C_0\in (K_D/\alpha, K_D]$:
\begin{eqnarray}\label{eq:perturbation-results}
    I_{\phi\phi} \approx \begin{cases} 
      \dfrac{N g^2(\alpha -1)^2 C_0^2  K_D^2\tilde{D} ^2}{8 \left(\alpha C_0-K_D\right)^4},  & \tilde{D}\ll1, \\
   \dfrac{Ng^2}{32}\left(1 - \dfrac{(\alpha C_0 - K_D)^2}{2\tilde{D}(\alpha-1)C_0K_D}   \right)^2,    & \tilde{D}\gg1.
   \end{cases}
\end{eqnarray}
As we expect, in the limit $\tilde{D}\to 0$, the Fisher information vanishes -- in this limit, the receptor binding is uniform across the cell. In the limit $\tilde{D} \gg 1$, we see that the Fisher information converges to the optimal value of $N g^2/32$. For both asymptotic limits, we note that at finite values of $\tilde{D}$, the Fisher information decreases as $C_0$ increases from $K_D/\alpha$ to $K_D$. {We believe this happens because when $G$ is high (smaller $C_0$), small variations in $G$  do not result in large changes in $K_\mathrm{eff}$, and in our model, a flatter profile of $K_\text{eff}(\theta)$ over the cell around a mean value of $K_\mathrm{eff} = C_0$ gives a higher $I_{\phi\phi}$.}
Fig.~\ref{fig:checking perturbation} shows that the numerical solution agrees with the perturbation results in both limits. %

\begin{figure}[tbp]
    \includegraphics[width=0.47\textwidth]{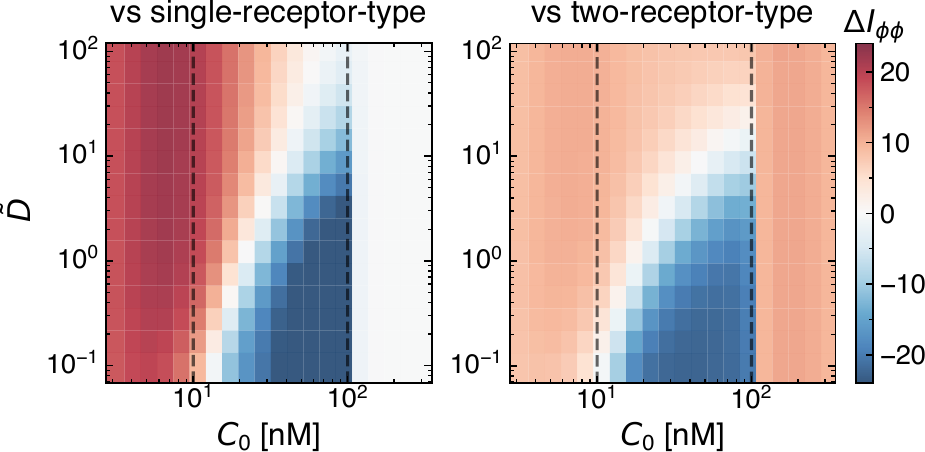}
    \caption{%
    Comparisons between Fisher information of the adaptation model, numerically solved, with the single receptor-type (left panel) and two receptor-type (right panel) models. The single-type model has dissociation constant $K_D$ and the two-type model has half its receptors with $K_D$ and the other half have $K_D/\alpha$. Black dashed lines show $C_0 =K_D/\alpha$ and $K_D$.  Dissociation constant $K_D=100\,$nM and $\alpha=10$.} 
    \label{fig:comparing-models} 
\end{figure}

Under what circumstances would our adaptation model be more accurate than a cell using a single receptor type \cite{hu2010mle}, or using two receptor types \cite{hopkins2020chemotaxis}, as we showed in Fig. \ref{fig:F_comparing}? We plot the difference in Fisher information between the adaptation model and these baseline models $\Delta I_{\phi\phi} = I_{\phi\phi}^\textrm{adapt}-I_{\phi\phi}^\textrm{baseline}$ in  Fig.~\ref{fig:comparing-models}. We see immediately that $\tilde{D}$ must be on the order of 10--100 in order for the adaptation model to outperform both models. Improvements over the two-receptor type model are less dramatic, but the difference in Fisher information $\Delta I_{\phi\phi}$ is positive over a larger region in the parameter space.

\begin{figure}[tbp]
    \includegraphics[width=0.47\textwidth]{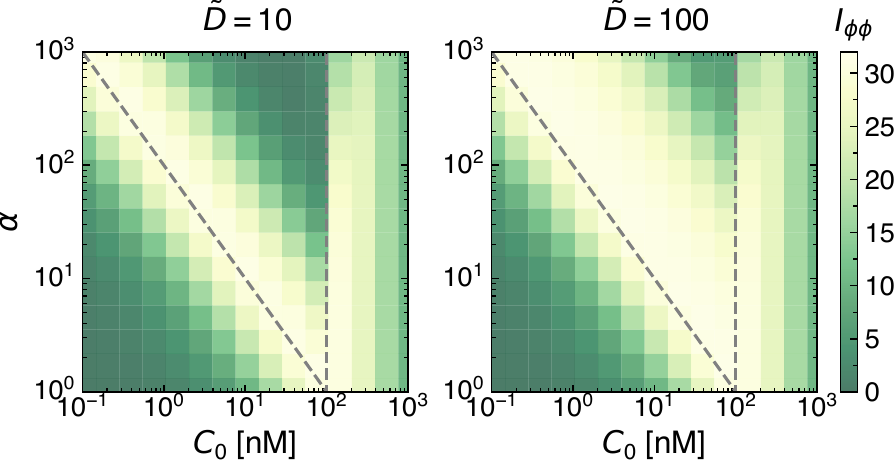}
    \caption{Fisher information for the adaptation model as a function of mean concentration $C_0$ and cooperativity $\alpha$ for two values of $\tilde{D}$. %
    A larger $\tilde{D}$ increases the range of mean concentrations over which the cell can sense accurately. Gray dashed lines indicate $K_D$ and $K_D/\alpha$, respectively; dissociation constant $K_D=100\,$nM.}
    \label{fig:C0_alpha_contour}
\end{figure}

We expect that the cooperativity $\alpha$ affects the accuracy of adaptation by determining the degree to which the allosteric protein affects ligand binding -- a naive guess would be that a larger cooperativity $\alpha$ would allow adaptation over a larger range of concentrations. However, this is only true in the limit $\tilde{D} \to \infty$. In Fig.~\ref{fig:C0_alpha_contour} we scan over values of $\alpha$ and the mean concentration $C_0$ at two different values of diffusion $\tilde{D} = 10$ and $\tilde{D} = 100$. We do see that at this highest value of $\tilde{D}$ that increasing $\alpha$ increases the range of concentrations $C_0$ where near-optimal sensing is possible. However, at lower $\tilde{D} = 10$, increasing $\alpha$ just shifts the region of accurate sensing around $K_D/\alpha$. %

\begin{figure}[tbp]
\includegraphics[width=0.43\textwidth]{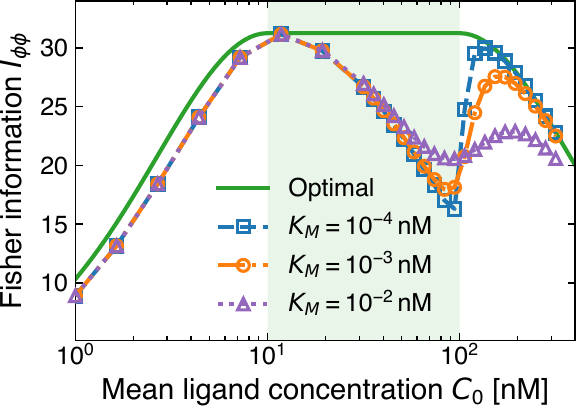}
\caption{Varying the Michaelis constant $K_M$, with $\tilde{D} = 10$. Right and left red dashed lines represent $C_0 = K_D/\alpha$ and $C_0 = K_D$, respectively. Note that the Fisher information jumps at $C_0 = K_D$ as the mechanism of adaptation does not function above this value of $C_0$.
Dissociation constant $K_D=100\,$nM and $\alpha=10$.
}
\label{fig:C0_KM}
\end{figure}

As discussed in Sec.~\ref{sec:model-MM}, increasing the Michaelis constant should affect gradient sensing accuracy at higher concentrations near $C_0 = K_D$. This can be seen in Fig.~\ref{fig:C0_KM}. When $K_M$ is sufficiently small, we observe a rapid increase in gradient sensing accuracy near $K_D$, after which the Fisher information seems to fall of like in the single-type model [Eq.~\eqref{eq:FI-single-receptor}]. {For larger $K_M$, however, the sharp jump is smoothed out.} This follows from the fact that smaller  $K_M$ makes it easier to maintain the limit $G,G^* \gg K_M$ where near-perfect adaptation is possible (Sec.~\ref{sec:model-MM}).

\subsection{Trade-offs in sensing: accuracy vs time to adapt}\label{sec:tradeoffs}

Our results show that the adaptation model is more successful when $\tilde{D} = DK_G /V_{\text{max}}R^2$ is large. This can occur either by increasing the diffusion coefficient of the allosteric protein or by decreasing the production rate $V_\textrm{max}/K_G$. However, because the diffusion coefficient of the protein cannot be increased arbitrarily, we think it is more plausible that $\tilde{D}$ can be increased by decreasing $V_\textrm{max}$. However, when the cell decreases $V_\textrm{max}$, this means that the conversion between $G$ and $G^*$ is slower -- so the time required for the cell to adapt to a new concentration will also slow.

\begin{figure*}[htbp]
    \includegraphics[width=0.99\linewidth]{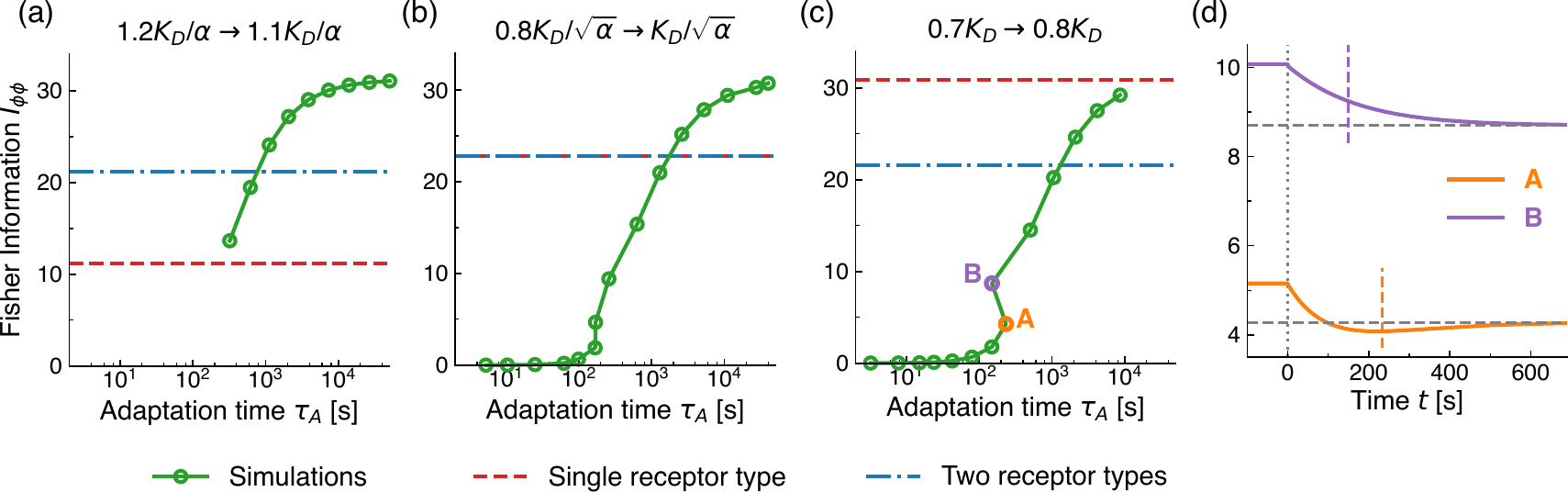}
    \caption{ Gradient sensing accuracy $I_{\phi\phi}$ must trade off with adaptation time $\tau_A$. Panels (a)--(c) show the simulated results for three different concentration jumps (labeled in title). Each data point corresponds to a simulation of Eq.~\eqref{eq:mm-kinetics-diff} with a different value of $V_\mathrm{max}$ {[$\tilde{D}$ ranges from $10^{-2}$ to $10^2$}. {We note that because we simulate at fixed $\tilde{D}$, the $\tau_A$ axis points do not match between the figure panels and some of the large-$\tilde{D}$ points are beyond the axis range in panel (a)].} 
    Panel (d) shows the time evolution of $I_{\phi\phi}(t)$ for two representative cases with subtle differences in dynamics, labeled \textbf{A} and \textbf{B} in panel (c). Concentration jump occurs at $t=0$; vertical dashed lines indicate the corresponding adaptation time $\tau_A$ calculated using Eq.~\eqref{eq:adap_t}.} 
    \label{fig:adapt-time-accuracy}
\end{figure*}

In Fig.~\ref{fig:adapt-time-accuracy}, we simulate gradient sensing and adaptation for a broad range of values of $\tilde{D}$. We treat the physical diffusion coefficient as fixed to $D = 0.1\ \textrm{\textmu}\mathrm{m^2/s}$ and cell size fixed to $R = 5 \ \textrm{\textmu} \textrm{m}$ Given these values and a value of $\tilde{D} = DK_G /V_{\text{max}}R^2$, this lets us set the characteristic timescale $K_G/V_\textrm{max}$ in real units. We define an adaptation time $\tau_A$ as the characteristic timescale for the change in the Fisher information in response to a jump in the concentration.
Assuming the jump occurs at time ${t}_0$ and the Fisher information relaxes to a steady-state value $I_{\phi\phi}^\textrm{ss}=\lim_{t\to\infty}I_{\phi\phi}({t})$, we define the adaptation time as:
\begin{equation}\label{eq:adap_t}
    \tau_A \equiv\frac{\displaystyle\int_{ t_0}^{\infty} ( t-{t}_0)\left|I_{\phi\phi}(t) - I_{\phi\phi}^\textrm{ss}\right|\mathrm{d}t}{\displaystyle\int_{ t_0}^{\infty}\left|I_{\phi\phi}(t) - I_{\phi\phi}^\textrm{ss}\right|\mathrm{d}t}.
\end{equation}
In Fig. \ref{fig:adapt-time-accuracy}, we compute this adaptation time for a broad range of $\tilde{D}$, and plot the steady-state Fisher information as a function of this adaptation time. We see generally that higher Fisher information requires longer adaptation times. We also compare the adaptation model to gradient sensing accuracy with  fixed receptor types (the blue dash-dot line indicates a 50-50 mix of $K_D$ and $K_D/\alpha$, the red dashed line a single receptor type of $K_D$). The adaptation time required for a particular Fisher information will depend on the final concentration $C_0$ (Fig. \ref{fig:adapt-time-accuracy}a-c), as we would expect based on the strong dependence of accuracy on $C_0$ in Fig. \ref{fig:checking perturbation}. We find that the adaptation time required for the adaptation model to beat the two-receptor-type model is generally of the order of a few hundred seconds. %
We note that interpreting the adaptation time may not be obvious because the Fisher information may not monotonically vary in adaptation (Fig. \ref{fig:adapt-time-accuracy}d).
Adaptation times around $\tau_A \sim 1000$ seconds might still be useful for slow cells like fibroblasts ~\cite{morrow2019integrating}, though these cells are also larger in size and would have correspondingly larger adaptation times. 
We show adaptation to relatively modest concentration changes in Fig. \ref{fig:adapt-time-accuracy}; we also test $\tau_A$ vs accuracy for a different concentration jump size to the same final concentrations (Fig.~\ref{fig:adapt_time_jump_size}), where we see that the size of the jump in concentration does not significantly affect the adaptation time.

\section{Discussion}\label{sec:discussion}

In this work, we have proposed a model of eukaryotic chemotaxis with near-perfect adaptation of receptor properties, allowing cells to sense gradients accurately over a wider range of average concentrations. We found that this approach's success depends on the relative rates of diffusion and modification of the allosteric protein, and therefore defined the dimensionless diffusion coefficient  $\tilde{D} = DK_G /V_{\text{max}}R^2$. Using a first-order perturbative approach, we derived analytical solutions for the Fisher information in the low and high $\tilde{D}$ limits. We evaluated the effectiveness of the model relative to other previously proposed models of chemotaxis with one/multiple receptor types. We find that adaptation works well for high $\tilde{D}$-- but this comes inevitably with a long adaptation time to a change in chemoattractant concentration.  %

Our model is intentionally simple, and  some of the assumptions made are relatively optimistic for the cell's ability to perform adaptation, and real cells may face significant additional challenges. For instance, we assume in our reaction kinetics [Eq.~\eqref{eq:mm-kinetics-nodiff}], that the inactivation rate does not depend on whether the allosteric protein is bound to the receptor. In thermal equilibrium, the rate of transition from $G \to G^*$ and its reverse are related to the change in free energy through detailed balance, and the free energy of the state will depend on whether G is bound. Our model is out of equilibrium, and adaptation therefore necessitates an energy cost for the cell, as has been previously proposed \cite{lan2012energy,tu2018adaptation}. As with the larger program of understanding energy costs for sensing \cite{ten2016fundamental}, these may further disfavor the adaptation model. 

Is adaptation plausible over biologically relevant timescales? For adaptation to have worthwhile results, the cell would need to have $\tilde{D}$ of order $10$, or an adaptation time of  on the order of 1000 seconds if $D = 0.1 \,\textrm{\textmu}\mathrm{m^2/s}$ (Fig.~\ref{fig:adapt-time-accuracy}).  Is $\tilde{D} \approx 10$ reasonable?  Diffusion coefficients of G-proteins are roughly of order of magnitude $0.1\,\textrm{\textmu}\mathrm{m^2/s}$ \cite{wang2008diffusion}, so the timescale of 1000 seconds is consistent with G proteins or other membrane-attached proteins as the allosteric protein. %
However, these values imply a $K_G/V_\textrm{max}$ of $\sim 1000$ s. This is much slower than typical timescales found in phosphorylation and methylation processes in cells \cite{blazek2015phosph}, but may nonetheless be plausible as such slow timescales have been found in some enzyme activity \cite{bareven2011enzymes}. Could cells function effectively if this adaptation time were so slow? We compare the adaptation timescale to how fast the cell's concentration environment changes. If a cell is traveling at a speed of $5\,\textrm{\textmu}\mathrm{m/min}$ in an environment $C(x) = C_0 e^{g x / L}$ and a $5\%$ gradient across the cell length  ($g = 0.05$),  then a cell must travel $\ln 2 / 0.05 \approx 14$ cell lengths to double the concentration $C_0$. This would take 28 minutes if the cell length is 10 $\textrm{\textmu}\mathrm{m}$. %
This suggests that -- in shallow gradients at least -- cells may see concentration changes that can be adapted to at relevant timescales. However, if there are rapidly-moving sources, concentration dynamics can be much faster and more complicated \cite{kashyap2024trade}, and we expect receptor adaptation at the timescale of hundreds or thousands of seconds to not be particularly useful. 

The adaptation mechanism we propose here may be more broadly relevant if the allosteric factor has a larger diffusion coefficient, e.g. if it is a primarily cytosolic protein that only occasionally binds to the membrane. {(In this case, we would have to more carefully treat a difference in diffusion coefficient between receptor-bound allosteric proteins, which will have low diffusion coefficients, and cytosolic. Within our current model the bound and unbound forms are lumped with a single effective diffusion coefficient.)}  More speculatively, some GPCRs respond to mechanical cues including membrane tension\cite{hardman2023membrane}, allowing the membrane itself to act as an ``allosteric'' factor \cite{sirbu2024cell,soubias2023physiological,levental2023regulation,fourel2025allosteric}. Tension could then serve as a global factor \cite{houk2012membrane} in place of an allosteric protein. However, we do note that tension propagation is not straightforward -- depending on cell type and on whether membrane or cortex tension is measured, very different extents of tension correlation and propagation can be found \cite{shi2018cell,shi2022membrane,houk2012membrane,de2023cell}.

A recent paper also proposed a related idea of using G-protein activation to improve gradient sensing, albeit with different underlying mechanisms. Ghose et al. \cite{ghose2025ratiometric} propose a model of ratiometric sensing where G-proteins serve the role of enhancing signal as well as temporally storing information on ligand from membrane receptors. We do not study the effect of time averaging here, instead focusing on the instantaneous state of whether receptors are bound or not. This is because time averaging in systems where receptor kinetics vary from receptor to receptor may be complicated and a naive average from integrating signals, as we would get from studying the signal-to-noise ratio of $G(t)$, can be very suboptimal \cite{hopkins2020chemotaxis}. We defer consideration of this point to future work. Surprisingly, given how different our mechanism is from that of the G protein model of \cite{ghose2025ratiometric}, Ghose et al. also find a tradeoff between adaptation time and accuracy, with adaptation times on the order of 50-100 seconds. We suspect the tradeoff between ability to adapt to new environments and accuracy in an individual environment is a universal tradeoff, potentially reflecting the time required to communicate information across the cell \cite{bryant2023physical}.

\begin{acknowledgments}
The authors acknowledge support from NIH Grant No.~R35GM142847 and NSF DMR 1945141. We thank Grace Luettgen and Emiliano Perez Ipi\~na for useful discussions and a close reading of the draft. 
\end{acknowledgments}

\appendix

\section{Solving the ternary complex model}\label{app:ternary-complex-sol}
We write the master equation corresponding to Figure~\ref{fig:ternary-complex}, using on/off rates that correspond to the given dissociation constants, as shown in Fig.~\ref{fig:ternary-complex-rates}. Without loss of generality, when the dissociation constant is reduced by a factor of $\alpha$, we choose the corresponding off rate to be reduced by a factor of $\alpha$. For the four states, we find the master equation:
\begin{figure}[b]
    \centering
    \includegraphics[width=0.7\linewidth]{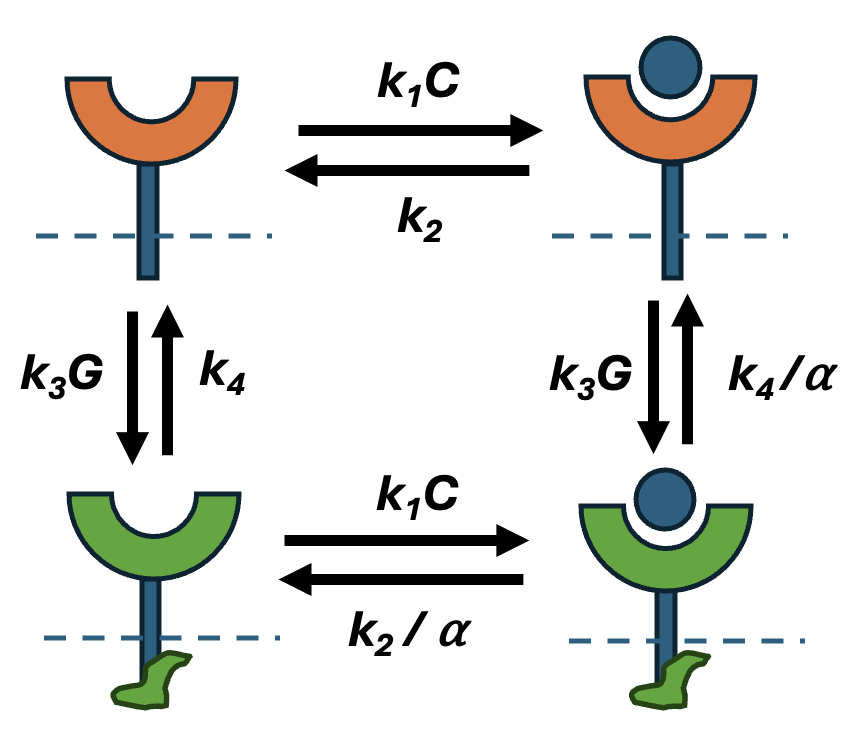}
    \caption{Ternary complex model diagram similar to Fig.~\ref{fig:ternary-complex}, with arrows labeled with forward/backward rates instead of dissociation constants. }
    \label{fig:ternary-complex-rates}
\end{figure}
\begin{subequations}
\begin{eqnarray}
    \frac{\mathrm{d}p^\circ}{\mathrm{d}t}&=& (-k_1C-k_3G)p^\circ + k_2p^\bullet + k_4p_G^{\circ},\label{eq:app-ternary-masters1}\\
    \frac{\mathrm{d}p^\bullet}{\mathrm{d}t}&=& k_1Cp^\circ + (-k_2-k_3G)p^\bullet + \frac{k_4}{\alpha}p^{\bullet}_G,\label{eq:app-ternary-masters2}\\
    \frac{\mathrm{d}p_G^{\circ}}{\mathrm{d}t}&=& k_3Gp^\circ + (-k_4-k_1C)p_G^{\circ} + \frac{k_2}{\alpha}p_G^{\bullet},\\
    \frac{\mathrm{d}p_G^{\bullet}}{\mathrm{d}t}&=& k_3Gp^\bullet + k_1Cp_G^{\circ}-\frac{k_2+k_4}{\alpha}p_G^{\bullet},\label{eq:app-ternary-masters4}
\end{eqnarray}
\end{subequations}
where $C$ and $G$ denote ligand and allosteric protein concentrations, the superscripts $\circ/\bullet$ denote ligand unbound/bound states,
and the subscript $G$ indicates that a allosteric protein molecule is bound to the receptor's cytosolic side.
We can take Eqs.~\eqref{eq:app-ternary-masters2}--\eqref{eq:app-ternary-masters4} in steady state, together with the constraint $p^\circ+p^\bullet+p_G^{\circ}+p_G^{\bullet}=1$:
\begin{equation*} \label{eq:app-ternary-steady}
    \begin{bmatrix}
        1&1&1&1\\
        k_1C &-k_2-k_3G &0 & k_4/\alpha\\
        k_3G&0&-k_4-k_1C&k_2/\alpha\\
        0&k_3G&k_1C&-\dfrac{k_2+k_4}{\alpha}
    \end{bmatrix}
    \begin{bmatrix}
        p^\circ_{\phantom{G}}\\
        p^\bullet_{\phantom{G}}\\
        p_G^{\circ}\\
        \vphantom{\dfrac{1}{1}}p_G^{\bullet}
    \end{bmatrix}
    = \begin{bmatrix}
        1\\0\\0\\\vphantom{\dfrac{1}{1}}0
    \end{bmatrix}.
\end{equation*}
To be consistent with the notation used in Fig.~\ref{fig:ternary-complex}, we define $K_D \equiv k_2/k_1$ and $K_G \equiv k_4/k_3$. Solving for $p^{\bullet}$ and $p_{G}^{\bullet}$ in this steady state, we obtain:
\begin{eqnarray*}
    p^\bullet &=& \frac{C}{C\left(1 +\frac{\alpha G}{K_G }\right) + K_D\left(1 + \frac{G}{K_G}\right)},\\
    p_G^{\bullet} &=& \frac{C}{C\left(1 + \frac{K_G}{\alpha G}\right) + \frac{K_D}{\alpha}\left(1 + \frac{K_G}{G}\right)}.
\end{eqnarray*}
Then, the total ligand-bound probability should be
\begin{equation}
     p_\mathrm{bound} = p^{\bullet} + p_G^{\bullet} =  \frac{C}{C + K_D\frac{G+K_G}{\alpha G+K_G}}.
\end{equation}

\section{Calculating the Fisher information from the bound probability}\label{app:fisher-information}
In this section, we generalize the results of \cite{hu2010mle,hu2011geometry} for the Fisher information of a gradient-sensing cell beyond simple ligand-receptor kinetics to any slowly-varying bound probability. If we have bound probability has the form 
\begin{equation}\label{eq:app-prob-bound}
    p_{\textrm{bound}} = \xi_0 + \xi_1g\cos(\theta-\phi),
\end{equation}
where $\xi_1g$ is small compared to $\xi_0$, we can, assuming Bernoulli statistics as in the main text and \cite{hu2010mle,hu2011geometry}, form a likelihood function $\mathcal{L}$, finding
\begin{widetext}
\begin{eqnarray}
&&\mathcal{L}\left(g, \phi \mid \{X_i\}\right) = \prod_{i=1}^N 
\left( \xi_0 + \xi_1 g \cos(\theta_i - \phi) \right)^{X_i} \times \left( 1 - \xi_0 - \xi_1 g \cos(\theta_i - \phi) \right)^{1 - X_i},\label{eq:app-likelihood}\\
&&\displaystyle\frac{\partial^2 \ln \mathcal{L}}{\partial \phi^2} = \sum_{i=1}^N\left[ X_i\left( -\frac{g^2 \xi_1^2 \sin^2(\theta_i - \varphi)}{(\xi_0 + g \xi_1 \cos(\theta_i - \varphi))^2} \right. - \frac{g \xi_1 \cos(\theta_i - \varphi)}{ \xi_0 + g\xi_1 \cos(\theta_i - \varphi)} \right) \nonumber \\
&&\qquad\qquad\qquad\quad\displaystyle-(1 - X_i) \left( \frac{g \xi_1 \cos(\theta_i - \varphi)}{1 - \xi_0 - g \xi_1 \cos(\theta_i - \varphi)} \left. + \frac{g^2 \xi_1^2 \sin^2(\theta_i - \varphi)}{(1- \xi_0 - g \xi_1 \cos(\theta_i - \varphi) )^2} \right)\right].\label{eq:app-2nd-deriv}
\end{eqnarray}
\end{widetext}
Taking the expected value, which replaces the Bernoulli variables $X_i$ with their associated probabilities, we can get the $(\phi,\phi)$ element of the Fisher information matrix to second order in $g$:
\begin{eqnarray}
    I_{\phi\phi} &=& \sum_{i=1}^N\left[\frac{g^2 \xi _1^2 \sin ^2(\theta_i -\varphi )}{\xi _0(1-\xi _0)} + \mathcal{O}(g^3)\right] \nonumber\\
    &\approx& \frac{N}{2\pi}\int_0^{2\pi} d\theta \left[\frac{g^2 \xi _1^2 \sin ^2(\theta -\varphi )}{\xi _0(1-\xi _0)} + \mathcal{O}(g^3)\right] \nonumber\\
    &=& \frac{Ng^2\xi_1^2}{2\xi_0(1-\xi_0)} + \mathcal{O}(g^3),\label{eq:I_phiphi}
\end{eqnarray}
where in the second equality we have approximated the discrete sum as an integral over the cell boundary in the limit of large receptor number $N$. 

{In fact, we can get the form of this Fisher information up to a geometric prefactor from a simple signal-to-noise calculation akin to \cite{rappel2008receptor,lakhani2017testing,lauffenburger1982influence}. If we simplify our geometry, treating the cell as a line, and have $N/2$ receptors on the front half of the cell and $N/2$ on the back half, the average difference in receptors bound between the back and the front is $\Delta n_b = \frac{N}{2} p_\textrm{bound}(\textrm{front})-\frac{N}{2} p_\textrm{bound}(\textrm{back}) = N \xi_1 g$. If the receptor binding events are independent, then the variance in $\Delta n_b$ is $\textrm{Var}(\Delta n_b) = \textrm{Var}(n_b(\textrm{front}) )+ \textrm{Var}(n_b(\textrm{back}))$. These variances are -- applying the central limit theorem to the N/2 Bernoulli events of a ligand binding -- $\textrm{Var}(\Delta n_b) = \frac{N}{2}  p_\textrm{bound}(\textrm{front})(1-p_\textrm{bound}(\textrm{front}) )+\frac{N}{2} p_\textrm{bound}(\textrm{back})(1-p_\textrm{bound}(\textrm{back}))$. To the lowest order in $g$, we can neglect the difference between the front and the back, and the probability of being bound is just $\xi_0$, so  $\textrm{Var}(\Delta n_b) = N \xi_0 (1-\xi_0)$. We thus find that a signal-to-noise ratio $\textrm{SNR} = \Delta n_b^2 / \textrm{Var}(\Delta n_b)$ is just $\textrm{SNR} = \frac{N g^2 \xi_1^2}{\xi_0 (1-\xi_0)}$. We can see that the Fisher information $I_{\phi\phi}$ is proportional to the signal-to-noise ratio.} 

\section{First-order perturbation to allosteric protein concentration}\label{app:perturbation-theory}
We present perturbative results for both the low- and high-diffusion regimes, which are first-order perturbations for $\tilde{D}\ll 1$ and $\tilde{D}\gg 1$, respectively.

\subsection{Low-diffusion regime}\label{sub:lowdiff}

If we assume that $G,G^* \gg K_M$, then the Michaelis terms in Eq.~\eqref{eq:mm-kinetics-dimless} become zero-order, and we find
\begin{equation}\label{eq:app-MM-limit}
    \frac{\partial \tilde G}{\partial \tilde t} = \tilde{D}\partial_\theta^2\tilde G - f_b(\tilde G,\theta)  + f_u(\tilde G,\theta),
\end{equation}
where $f_b$ is the local bound fraction of receptors. %
We note that this approximation only holds in the range $K_D/\alpha \leq C_0 \leq K_D$, as outside that range, we cannot self-consistently have both $G$ and $G^*$ large relative to $K_M$. In the limit of $\tilde{D} = 0$, the bound fraction will adapt to exactly 1/2 everywhere. We can consider a first-order perturbation from the zero-diffusion case $\tilde{D}=0$:
\begin{equation}
    \tilde G = \tilde G_{1/2}(\theta) + \epsilon \tilde G_1(\theta)+\mathcal{O}(\epsilon^2),
\end{equation}
where $\epsilon\equiv \tilde{D}$, and $\tilde G_{1/2}(\theta)$ %
is such that $f_b\bm{(}\tilde G_{1/2}(\theta), \theta\bm{)}=1/2$, which is what we see when $\tilde{D}=0$. The term $f_b(\tilde G, \theta)$ then becomes $f_b\approx 1/2  + \epsilon \tilde G_1\left.({\partial f_b}/{\partial \tilde G})\right|_{\tilde G=\tilde G_{1/2}}$ and similarly $f_u(\tilde G,\theta)\approx  1/2  - \epsilon \tilde G_1\left.({\partial f_b}/{\partial \tilde G})\right|_{\tilde G=\tilde G_{1/2}}$.
In steady state ($\partial \tilde G/\partial \tilde t = 0$), to first order in $\epsilon$, we have from Eq. \ref{eq:app-MM-limit}, 
\begin{equation}
    \epsilon \partial_\theta^2 \tilde G_{1/2} - 2\epsilon \tilde G_1\left.\frac{\partial f_b}{\partial \tilde G}\right|_{\tilde G=\tilde G_{1/2}} = 0.
\end{equation}
from which we obtain
\begin{equation}\label{eq:first-ord}
    \tilde G_1 = {\partial_\theta^2\tilde G_{1/2}}\Big/{2\left.\frac{\partial f_b}{\partial \tilde G}\right|_{\tilde G_{1/2}}}.
\end{equation}
We can compute both these terms. We can find
$$\tilde G_{1/2}(\theta) = \frac{K_D - C_0\exp\left(\frac{g}{2}\cos(\theta-\phi)\right)}{\alpha C_0\exp\left(\frac{g}{2}\cos(\theta-\phi)\right) - K_D},$$
from solving $C(\theta)/(C(\theta)+K_\textrm{eff}) = 1/2$. Taking only terms up to first order in $g$ in this formula, we get
\begin{equation}
    \partial_\theta^2\tilde G_{1/2} = \frac{(\alpha-1)C_0K_D\cos(\theta-\phi)}{2(\alpha C_0 - K_D)^2}g.
\end{equation}
The local average fraction of bound receptors is
$f_b = {C}/{(C+K_\textrm{eff})}$,
where the dimensionless $K_\textrm{eff} = K_D (\tilde{G} + 1)/(\alpha \tilde{G} + 1)$.
Evaluating the other term in Eq.~\eqref{eq:first-ord}:
\begin{equation}
    \frac{\partial f_b}{\partial \tilde G} = \frac{(\alpha-1)CK_D}{(K_D(\tilde G+1) + C(\alpha G + 1))^2},
\end{equation}
we obtain
\begin{equation}
     \left.\frac{\partial f_b}{\partial \tilde G}\right|_{G_{1/2}} = \frac{(\alpha C- K_D)^2}{4(\alpha-1)CK_D}, 
\end{equation}
where $C$ is given in Eq.~\eqref{eq:concentration-profile}.

Finally, we have, to first order in $g$,
\begin{equation}
    \tilde G_1(\theta) \approx \frac{(\alpha-1)^2C_0^2K_D^2\cos(\theta-\phi)}{(\alpha C_0 - K_D)^4}g.
\end{equation}
Notice the $\cos(\theta-\phi)$ term. If we look at $\tilde G_{1/2}(\theta)$, it seems to follow the form $\Gamma_0 + \Gamma_1g\cos(\theta-\phi)$ (at least, to first order in $g$). We can substitute the first-order expansion of $\tilde G_{1/2}$ in $g$, then, into the final expression for $G$:
\begin{widetext}
\begin{eqnarray}
    \tilde G(\theta) &= &\frac{K_D-C_0}{\alpha C_0-K_D} + g\cos(\theta-\phi)\left(-\frac{C_0K_D(\alpha-1)}{2(\alpha C_0-K_D)^2} +\epsilon\frac{(\alpha-1)^2C_0^2K_D^2}{(\alpha C_0 - K_D)^4} \right)\nonumber\\
    &=& \tilde G_\textrm{opt} + g\cos(\theta-\phi)\frac{C_0K_D(\alpha-1)}{2(\alpha C_0-K_D)^2}\times\left(-1 + 2\epsilon\frac{(\alpha-1)C_0K_D}{(\alpha C_0 - K_D)^2} \right)\nonumber\\
    &=& \tilde G_\textrm{opt} + g\cos(\theta-\phi)\dfrac{C_0K_D(\alpha-1)}{2(\alpha C_0-K_D)^2}\times\left(-1 + 2\tilde{D}\dfrac{(\alpha-1)C_0K_D}{(\alpha C_0 - K_D)^2} \right),\label{eq:low-pert-gprotein}
\end{eqnarray}
\end{widetext}
where $\tilde G_{\textrm{opt}} = {(K_D-C_0)}/{(\alpha C_0-K_D)}$ is the (uniform) allosteric protein concentration that maximizes the Fisher information when $K_{\textrm{eff}}$ is constant across the cell. Then, plugging this into Eq.~\eqref{eq:p_bound}, we get
\begin{equation*}
    p_\text{bound} = \frac{1}{2} + \frac{\tilde D(\alpha -1) C_0 K_D }{4  \left(K_D-\alpha  C_0\right)^2} g\cos (\theta -\phi) + \mathcal{O}(g^2).
\end{equation*}
 Then, using Eq.~\eqref{eq:I_phiphi} with $\xi_0 = 1/2$ and $\xi_1={D(\alpha -1) C_0 K_D }/{4  \left(K_D-\alpha  C_0\right)^2}$,  we obtain
\begin{equation}\label{eq:low-pert-result}
    I_{\phi\phi} \approx \frac{Ng^2(\alpha -1)^2 C_0^2 \tilde{D}^2 K_D^2}{8\left(\alpha C_0 - K_D\right)^4} + \mathcal{O}(g^3).
\end{equation}

\subsection{High-diffusion regime}
Similar to the low $\tilde{D}$ case, we now assume the final solution for $\tilde{D}\gg 1$ includes a small first-order correction $\epsilon G_1$ to $G_\textrm{opt}$, but with $\epsilon\equiv 1/\tilde{D}$:
\begin{equation}\label{app:high-diff-1st-order}
    \tilde G = \tilde G_{\textrm{opt}} + \epsilon \tilde G_1(\theta) + \mathcal{O}(\epsilon^2),
\end{equation}
We note here the optimal $G$ (high-$\tilde{D}$ limit) is constant. Again, with $G, G^*\gg K_M$, Eq.~\eqref{eq:app-MM-limit} gives, in the steady state:
\begin{eqnarray}\label{eq:app-taylor-expand-high-diff}
    0&= & 1 - 2f_b(\tilde G,\theta) + \frac{1}{\epsilon}\partial_\theta^2(\epsilon \tilde G_1) + \mathcal{O}(\epsilon)\nonumber\\
    &=& 1 - 2f_b(\tilde G_{\textrm{opt}},\theta)  +  \partial_\theta^2\tilde G_1 + \mathcal{O}(\epsilon)
\end{eqnarray}
where in the first equality we have used $f_u = 1-f_b$. Note that $f_b = C/(C+K_{\textrm{eff}})$ depends on the concentration profile, so we can expand it to first order in $g$ as $f_b(\tilde G)\approx ({1}/{2})+({g}/{8})\cos(\theta-\phi)$.
Then, comparing the terms of order $\epsilon^0$ in Eq.~\eqref{eq:app-taylor-expand-high-diff}:
\begin{equation}
    \partial_\theta^2 \tilde G_1 \approx \frac{g}{4} \cos(\theta-\phi),
\end{equation}
which can be simply solved via ansatz to give
\begin{equation}\label{eq:app-high-pert-correction}
    \tilde G_1(\theta) = -\frac{g}{4} \cos(\theta-\phi).
\end{equation}
Finally, the correction is
\begin{equation}\label{eq:app-high-pert-gprotein}
    \tilde G(\theta) \approx \tilde G_\textrm{opt} - \frac{g}{4\tilde{D}}\cos(\theta-\phi).
\end{equation}
Following the same steps as Appendix~\ref{sub:lowdiff}, we can get from \eqref{eq:FI-pbound}:
\begin{equation}
    p_\mathrm{bound} = \frac{1}{2} + \frac{1}{8} g \cos (\theta -\varphi ) \left(1-\frac{\left(K_D-\alpha  C_0\right)^2}{2\tilde D(\alpha -1) C_0 K_D}\right) +\mathcal{O}(g^2)
\end{equation}
Then again, with $\xi_0 = 1/2$ and $\xi_1 = \frac{1}{8}\left(1-\frac{\left(K_D-\alpha  C_0\right)^2}{2\tilde D(\alpha -1) C_0 K_D}\right)$, we arrive at the Fisher information:
\begin{equation}\label{eq:app-high-pert-result}
    I_{\phi\phi} = \frac{Ng^2}{32}\left(1 - \frac{(\alpha C_0 - K_D)^2}{2\tilde{D}(\alpha-1)C_0K_D}\right)^2.
\end{equation}
We can see this more easily as a small correction to the optimal Fisher information $Ng^2/32$.

\section{Simulation details}

All simulations of the second-order PDEs like Eq. \eqref{eq:mm-kinetics-diff} are executed by Euler's method of numerical integration using the forward-time centered-space method. We update \eqref{eq:mm-kinetics-dimless} as follows:
\begin{widetext}
    \begin{eqnarray}\label{app:simulated-mm}
     \tilde G(\theta,\tilde t+\Delta \tilde t) &=&  \tilde G(\theta,\tilde t)+\frac{{\tilde D}\Delta \tilde t}{(\Delta \theta)^2}\left[\tilde G(\tilde t, \theta + \Delta \theta)-2 \tilde G(\tilde t,\theta)+ \tilde G(\tilde t, \theta-\Delta \theta)\right]\nonumber\\
    &&\phantom{\tilde G(\theta,t)}+ \Delta \tilde t\left[f_b(\tilde t,\theta)\frac{ \tilde G(\tilde t,\theta)}{ \tilde G(\tilde t,\theta)+\tilde K_M} - f_u(\tilde t,\theta)\frac{ \tilde G^*(\tilde t,\theta)}{ G^*(\tilde t,\theta)+\tilde K_M}\right].
\end{eqnarray}
\end{widetext}
We update \eqref{app:simulated-mm} for 40 points on the cell membrane ($\Delta \theta = \pi/20 \approx 0.157$)  until steady state and use Eq. \eqref{eq:FI-pbound} after fitting the bound probability to Eq. \eqref{eq:app-prob-bound}. Note that in doing so, we are assuming that the time scale associated with receptor-ligand binding is much shorter than the time scale associated with the allosteric protein's activation/inactivation and diffusion. 

For some parameter values, including for low values of $\tilde D$ and $C_0$, reaching steady state may require up to $\sim 3.3\times10^2$ in unitless time.  In these cases, we can both increase $\Delta t$ and the maximum number of time steps, the former of which can be afforded due to the low diffusion coefficient and the (approximate) stability condition $\frac{D\Delta t}{(\Delta x)^2}\lesssim \frac{1}{2}$. Unless we are sweeping over any of the parameters listed below, they remain at the values listed in Table~\ref{tab:param}.
For the adaptation time calculations in Fig.~\ref{fig:adapt-time-accuracy}, we run the simulation for a total time of $\tilde{T}=1000$ for the $1.2K_D/\alpha\to1.1K_D/\alpha$ concentration jump, and $\tilde{T}=200$ for the other two. The concentration jump occurs at time $\tilde{t}_0=\tilde{T}/3$, allowing the system to reach steady state at both the initial and final concentrations. We then define the unitless adaptation time $\tilde{\tau}_A$:
\begin{equation}
    \tilde{\tau}_A \equiv \frac{\displaystyle\int_{\tilde{t}_0}^{\tilde T} (\tilde t - \tilde t_0)\left|I_{\phi\phi}(\tilde{t})-I_{\phi\phi}(\tilde{T})\right|\mathrm{d}\tilde t}{\displaystyle\int_{\tilde t_0}^{\tilde T} \left|I_{\phi\phi}(\tilde{t})-I_{\phi\phi}(\tilde{T})\right|\mathrm{d}\tilde t},
\end{equation}
thus the physical $\tau_A$ can be computed from the unitless form as $\tau_A = \tilde{\tau}_A \times K_G/V_\textrm{max}$.%

\begin{table*}
\caption{\label{tab:param}%
Table of simulation parameters. \footnote{These parameters are used throughout the paper; any deviation from them is explicitly noted. Description of the parameter and their dimensions are given along with their value in ``dimensionless'' units---for example, the values of $\tilde{G}$, $\tilde{K}_M$ in Eq.~\eqref{eq:mm-kinetics-dimless}.}}
\begin{ruledtabular}
\begin{tabular}{llccc}
Parameter & \qquad Description & Unit & Dimensionless value & \\\hline
$g$ & Gradient steepness & $1$ &  $0.05$ & \\
$C_0$  & Mean concentration of ligands & nM & \\
$N$ & Number of receptors & $1$ & $4\times 10^5$\\
$\Delta \theta$ & Angular spacing between simulated G-protein sites  & 1 & $\pi/20$\\
$K_D$ & Dissociation constant with ligand & nM & $100$ & \\
$\alpha$ & Receptor cooperativity & $1$ & $10$ & \\
$K_G$ & Dissociation constant with allosteric factor & nM & $1$ & \\
$K_M$ & Michaelis constant & nM & $10^{-4}\ (=K_M/K_G)$ & \\
$V_\textrm{max}$ & Limiting rate of allosteric protein activation & nM/s & 1\\
$D$ & Diffusion coefficient of allosteric proteins & $\textrm{\textmu}\mathrm{m^2/s}$ &\\
$\Delta t$ & Simulation time step & s & $10^{-5}\ (=\Delta t\frac{V_\mathrm{max}}{K_G})$& \\
$G_\mathrm{tot}$ & Total allosteric protein concentration & nM & $50\ (=G_\mathrm{tot}/K_G)$
\end{tabular}
\end{ruledtabular}
\end{table*}

\counterwithin*{figure}{part}
\stepcounter{part}
\renewcommand{\thefigure}{S\arabic{figure}}

\begin{figure*}[htbp]
    \includegraphics[width=0.85\textwidth]{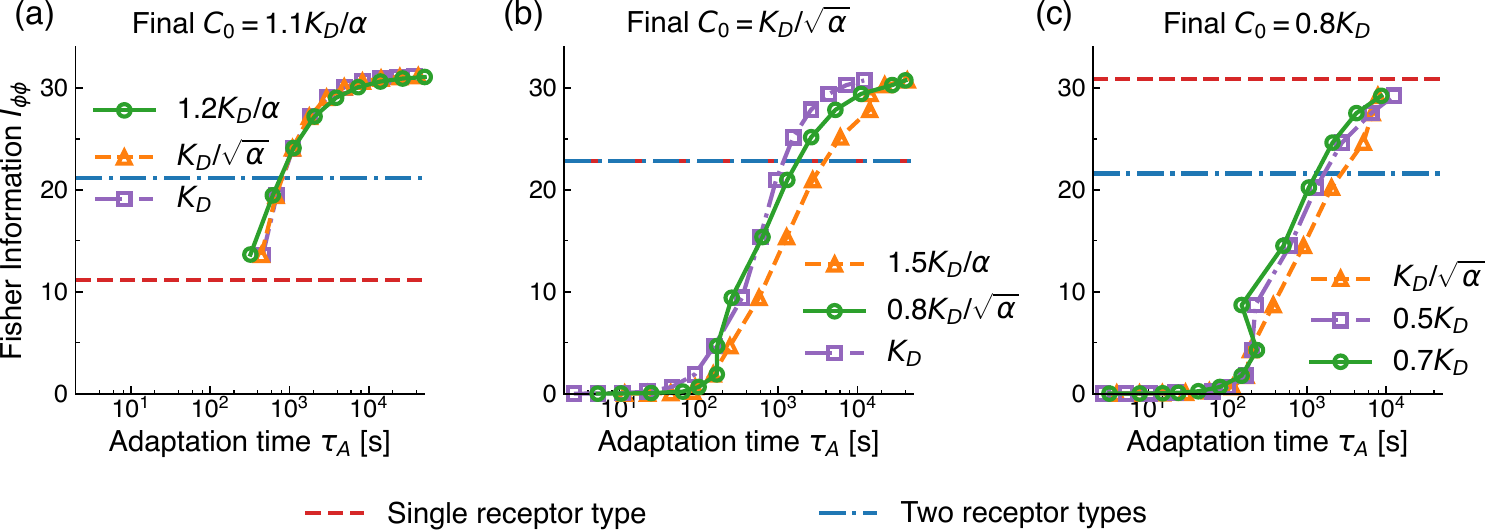}
    \caption{Adaptation time $\tau_A$ for different concentration jump sizes to the same final concentrations (labeled in title). Each panel shows the simulation results (empty symbols) for three different initial concentrations. Some of the large-$\tilde{D}$ points are beyond the axis range in panel (a).} 
    \label{fig:adapt_time_jump_size}
\end{figure*}

\clearpage

\bibliography{main}%

\end{document}